%% file: main.tex
\documentclass[letterpaper,twocolumn,10pt]{article}

\usepackage[utf8]{inputenc}
\usepackage[english]{babel}
\usepackage[T1]{fontenc}
\usepackage{lmodern}
\usepackage[svgnames,x11names]{xcolor}
\usepackage{hyphenat}

\usepackage{usenix2019_v3}

\usepackage{multirow}
\usepackage{tabularx}

\usepackage[backend=biber,
bibstyle=numeric,
citestyle=numeric-comp,
sortcites=true,
maxbibnames=10,
]{biblatex}
\renewcommand*{\bibfont}{\footnotesize}

\usepackage[]{graphicx}
\usepackage{xspace}
\usepackage{amsmath}
\usepackage{amssymb}
\usepackage{amsfonts}
\usepackage{amsfonts}
\usepackage{amsthm}
\usepackage{tabulary}
\usepackage{pifont}
\PassOptionsToPackage{hyphens}{url}
\usepackage{float}
\usepackage{adjustbox}
\usepackage{booktabs}
\usepackage{wasysym}
\usepackage{xstring}
\usepackage{soul}
\usepackage{enumerate}
\usepackage[olditem,oldenum]{paralist}
\usepackage{hyperref}
\usepackage{cleveref}
\usepackage{csquotes}
\usepackage[binary-units]{siunitx}
\usepackage{balance}

\hypersetup{
  colorlinks,
  linkcolor={black},
  citecolor={red!70!black},
  urlcolor={blue!70!black}
}

\usepackage[disable]{todonotes}

\input{inc/lstconfig}

\input{inc/circled}
\input{inc/macros}
\makeatletter%
\if@todonotes@disabled%
\else%
\fi
\makeatother

\begin{document}

\title{%
\toolname: Discovery and Exploitation of Memory Corruption\\Vulnerabilities in SGX Enclaves}
\author{Tobias Cloosters, Michael Rodler, Lucas Davi \\
    University of Duisburg-Essen, Germany\\
    \textit{\{tobias.cloosters, michael.rodler, lucas.davi\}@uni-due.de} \\
}
\date{}

\maketitle

\begin{abstract}

\input{sections/abstract.tex}
\end{abstract}

\section{Introduction}%
\label{sec:intro}

\input{sections/introduction.tex}

\section{Memory Corruption in SGX}%
\label{sec:problem}

\input{sections/problem-sgx-memory-corruption.tex}

\section{SGX Preliminaries}%
\label{sec:background}

\input{sections/background.tex}

\section{\toolname\ Symbolic Enclave Analyzer}%
\label{sec:architecture}

\input{sections/symex-implementation.tex}

\section{Enclave Analysis Results}%
\label{sec:evaluation}

\input{sections/results.tex}

\section{Performance and Accuracy}

\input{sections/performance.tex}

\section{Discussion}%
\label{sec:discussion}

\input{sections/discussion.tex}

\section{Related Work}%
\label{sec:relatedwork}
\input{sections/relatedwork.tex}

\section{Conclusion}%
\label{sec:conclusion}

\input{sections/conclusion.tex}

\section*{Acknowledgment}

We would like to thank the affected vendors, Intel, Baidu, WolfSSL, \Synaptics, \Goodix, and the enclave developers for promptly acting upon our reports and developing patches.
Furthermore, we especially thank our shepherd, Nathan Dautenhahn, for helping us to improve this work.
Funded by the Deutsche Forschungsgemeinschaft (DFG, German Research Foundation) under Germany's Excellence Strategy - EXC 2092 CASA - 390781972 and under SFB 1119 – 236615297.

{\footnotesize
	\printbibliography
}

\end{document}

%% file: inc/lstconfig.tex
\usepackage{listings}

\definecolor{lstgreen}{rgb}{0,0.6,0}
\lstdefinestyle{plain}{%
  numbers=none,
  frame=none,
  xleftmargin=1pt,
  xrightmargin=1pt,
}

\lstdefinelanguage{EDL}[ISO]{C++}{%
  morekeywords=[2]{user_check,in,out,size,string}%
}%

\lstdefinelanguage{Rust}{%
  morekeywords={as,break,const,continue,crate,else,enum,extern,false,
    fn,for,if,impl,in,let,loop,match,mod,move,mut,pub,ref,return,Self,
    self,static,struct,super,trait,true,type,unsafe,use,where,while,
    abstract,alignof,become,box,do,final,macro,offsetof,override,priv,
    proc,pure,sizeof,typeof,unsized,virtual,yield},
  morekeywords=[2]{isize,usize,char,bool,str,String,u8,u16,u32,u64,u128,i8,i16,i32,i64,i128,f32,f64},
  sensitive=true,
  morecomment=[l]{//},
  morecomment=[l]{///}, %
  morestring=[b]{"},
}%

%% file: inc/circled.tex
\usepackage{tikz}
\usetikzlibrary{shapes,arrows,positioning,shapes.multipart}

\newcommand{\circleone}{\ding{192}\xspace}

\newcommand{\circletwo}{\ding{193}\xspace}
\newcommand{\circlethree}{\ding{194}\xspace}
\newcommand{\circlefour}{\ding{195}\xspace}
\newcommand{\circlefive}{\ding{196}\xspace}

%% file: inc/macros.tex
\newcommand{\intel}{Intel\xspace}
\newcommand{\sgxsdk}{\intel\ SGX SDK\xspace}

\newcommand{\p}[1]{\SI{#1}{\percent}}
\newcommand{\s}[1]{\SI{#1}{\second}}
\newcommand{\gb}[1]{\SI{#1}{\giga\byte}}

\let\underscore\_
\renewcommand{\_}{$\underscore$\allowbreak{}}%

\newcommand{\sym}[1]{\texttt{#1}\xspace}

\newcommand\eenter{\emph{EENTER}\xspace}
\newcommand\eexit{\emph{EEXIT}\xspace}

\newcommand\narrowellipsis[1][-2pt]{.\hspace{#1}.\hspace{#1}.}
\newlength{\nowidthtmp}\newcommand{\nowidth}[1]{#1{\settowidth{\nowidthtmp}{#1}\hspace{-\nowidthtmp}}}

\newcommand{\toolname}{\textsc{TeeRex}\xspace}
\newcommand{\longtoolname}{{Trusted Enclave Ecall Runtime EXploiter}\xspace}

\newcommand{\scname}[1]{{\textsc{#1}}{}\xspace}

\newcommand{\angr}{\scname{Angr}{}\xspace}
\newcommand{\ecalls}{ECALLs\xspace}
\newcommand{\ecall}{ECALL\xspace}
\newcommand{\ocalls}{OCALLs\xspace}
\newcommand{\ocall}{OCALL\xspace}

\newcommand{\hex}[1]{\(\mathtt{#1}\)}
\newcommand{\gitcommit}[1]{\scriptsize\(\mathtt{\StrLeft{#1}{13}}\)}
\newcommand{\version}[1]{#1}

\renewcommand{\paragraph}[1]{\medskip\noindent\textbf{#1}\hspace{1ex minus .1ex}}

\newcommand{\citewauthor}[1]{\citeauthor*{#1}~\cite{#1}}

\makeatletter
\newcommand*{\rom}[1]{\expandafter\@slowromancap\romannumeral #1@}
\makeatother

\newcommand{\pattern}[1]{\paragraph{#1.}\xspace}
\newcommand{\component}[1]{\paragraph{#1:}}

\newcounter{challenge}
\newcommand\challenge[1]{\stepcounter{challenge}\paragraph{C\thechallenge: #1.}}

\newcommand{\Synaptics}{Synaptics\xspace}
\newcommand{\Goodix}{Goodix\xspace}
\newcommand{\Lenovo}{Lenovo\xspace}
\newcommand{\Dell}{Dell\xspace}

\newcommand{\enclaveIntelGMP}{Intel GMP Example\xspace}
\newcommand{\enclaveRustTLS}{Rust SGX SDK's tlsclient\xspace}
\newcommand{\enclaveTalos}{TaLoS\xspace}
\newcommand{\enclaveWolfSSL}{WolfSSL Example Enclave\xspace}
\newcommand{\enclaveSynaTEE}{Synaptics SynaTEE Driver\xspace}
\newcommand{\enclaveSynaTEEversion}{\version{5.2.3535.26}\xspace}
\newcommand{\enclaveGoodix}{Goodix Fingerprint Driver\xspace}
\newcommand{\enclaveGoodixVersion}{\version{2.1.32.200}\xspace}
\newcommand{\enclaveSignal}{SignalApp Contact Discovery\xspace}

\newcommand{\titlevulnPtrInDS}{P1: Passing Data-Structures with Pointers}
\newcommand{\titlevulnResource}{P2: Returning pointers to enclave memory}
\newcommand{\titlevulnOverlap}{P3: Pointers to Overlapping Memory}
\newcommand{\titlevulnNull}{P4: NULL-Pointer Dereferences}
\newcommand{\titlevulnTOCTOU}{P5: Time-of-Check Time-of-Use}

\newcommand{\vulnPtrInDS}{\emph{Passing Data-Structures with Pointers (P1)}\xspace}
\newcommand{\vulnResource}{\emph{Returning pointers to enclave memory (P2)}\xspace}
\newcommand{\vulnOverlap}{\emph{Pointers to Overlapping Memory (P3)}\xspace}
\newcommand{\vulnNull}{\emph{NULL-Pointer Dereferences (P4)}\xspace}
\newcommand{\vulnTOCTOU}{\emph{Time-of-Check Time-of-Use (P5)}\xspace}

%% file: sections/abstract.tex
\intel's Software Guard Extensions (SGX) introduced new instructions to switch the processor to \emph{enclave mode} which protects it from introspection.
While the enclave mode strongly protects the memory and the state of the processor, it cannot withstand memory corruption errors inside the enclave code.
In this paper, we show that the attack surface of SGX enclaves provides new challenges for enclave developers as exploitable memory corruption vulnerabilities are easily introduced into enclave code.
We develop \toolname\ to automatically analyze enclave binary code for vulnerabilities introduced at the host-to-enclave boundary by means of symbolic execution.
Our evaluation on public enclave binaries reveal that many of them suffer from memory corruption errors allowing an attacker to corrupt function pointers or perform arbitrary memory writes.
As we will show, \toolname\ features a specifically tailored framework for SGX enclaves that allows simple proof-of-concept exploit construction to assess the discovered vulnerabilities.
Our findings reveal vulnerabilities in multiple enclaves, including enclaves developed by Intel, Baidu, and WolfSSL, as well as biometric fingerprint software deployed on popular laptop brands.

%% file: sections/introduction.tex
Intel recently introduced a sophisticated trusted execution environment (TEE) called Software Guard Extensions (SGX)~\cite{Hoekstra2013-hasp-sgx,McKeen2013-hasp-sgx-apps,sgxref}.
SGX allows application developers to create so-called \emph{enclaves} to encapsulate sensitive application code and data inside a TEE that is completely isolated from other applications, operating systems, and hypervisors.
The only trusted component in the SGX setting is the Intel CPU itself.
Most prominently, SGX features confidentiality and integrity protection for any data that is written to its main memory.
In addition, SGX implements well-known Trusted Computing concepts such as data binding and sealing as well as remote attestation, i.e., ensuring the remote SGX enclave is in a trustworthy state.
Putting all these features together, this allows a user to establish a secure channel directly to the SGX enclave (which potentially runs in an untrusted cloud environment) and perform remote attestation to ensure the integrity of the remote SGX hardware and enclave.
That said, SGX is a strong isolation mechanism for sensitive data (e.g., personal information or cryptographic keys) as well as security-critical code (e.g., for the sake of intellectual property protection).
It also found its way into commercial applications, e.g., fingerprint sensor software (\Cref{sec:evaluation}), DRM protection~\cite{cyberlink_powerdvd}, and privacy-preserving applications like Signal~\cite{signal-enclave}.
As such a promising technology, SGX has been used and targeted extensively in previous research.
Many projects propose to utilize SGX for enhanced security guarantees, e.g., processing private data in public clouds~\cite{schuster2015vc3,baumann2014shielding}.

From its infancy, it was clear that SGX cannot withstand all flavors of attacks~\cite{sgx-side-channels}.
In particular, SGX cannot protect against two classes of attacks: (1)~side-channel attacks and (2)~memory corruption attacks inside the enclave.
The former attack technique exploits shared resources (e.g., cache) to steal secret information from within an enclave.
This line of research has become a very active research field~\cite{Xu2015-controlledchannel,VanBulck2017-stealthyptattacks}.
Especially micro-architectural side-channels have been shown to be effective for attacking SGX enclaves due to the shared micro-architectural state of enclaves and untrusted code~\cite{VanBulck2018-foreshadow}.

To our surprise, memory corruption attacks have been rarely investigated in the context of SGX.
These attacks exploit programming errors (e.g., a buffer overflow) allowing an attacker to take over the enclave, hijacking the enclave's control-flow, and perform code-reuse attacks such as return-oriented programming (ROP)~\cite{shacham2007rop}.
Further, the attacker can also exploit these errors to corrupt enclave data variables and pointers to launch data-oriented attacks such as information leaks or data-oriented programming (DOP)~\cite{Hu2016-dop}.
Prior research studied the applicability of offensive and defensive techniques against memory corruption exploits.
For instance, \citewauthor{lee2017darkrop} presented \scname{DarkROP}, a code-reuse attack technique, which shows that the enclave code must not be known to an attacker to successfully launch ROP attacks against the enclave.
\citewauthor{Biondo2018-dilemma} showed that it is easily possible to launch powerful code-reuse attacks due to particularities of the Intel SGX SDK bypassing existing ASLR defenses such as SGX-Shield~\cite{seo2017sgxshield}.

However, prior research on memory corruption attacks \emph{always assumed the existence of memory errors, but did not investigate whether or to which extent such errors exist in real-world enclaves}.
Due to the rather slow adoption of the SGX technology, this is not an easy question to answer.
Ideally, SGX enclaves contain only a minimal amount of code, which can be manually audited or even formally verified to not contain any programming mistakes.
However, in our experience, legacy code bases are often ported to SGX enclaves.
These ports are often not revised to handle the specialties of SGX enclaves and inherit security vulnerabilities from the legacy code base or introduce new security vulnerabilities particular to SGX enclaves.
This is similar for newly written SGX code by developers not familiar with the peculiarities of SGX.

One common aspect of all SGX enclaves is that they always link to an untrusted \emph{host application}.
The host application loads an SGX enclave into its address space as it would do in case of a shared library.
Indeed, the \sgxsdk offers a C-function like interface allowing bidirectional communication from the host application to the enclave.
This interface is highly critical as invalid input may lead to a privilege escalation attack.
As shown by prior research in the context of other privilege separation technologies, this is especially true when software is partitioned into privilege levels~\cite{checkoway2013iago,Hu2015-privsepvulns}.
That said, whenever an enclave is called, it must take special care to validate any input, particularly when the input contains code or data pointers.

\paragraph{Contributions.}
In this paper, we demonstrate that the attack surface of SGX enclaves provides new challenges for enclave developers as exploitable memory corruption vulnerabilities are easily introduced into enclave code due to a combination of the unique threat model of SGX enclaves and the current prevalent programming model for SGX (i.e., the \sgxsdk).
We introduce the first SGX vulnerability analysis framework, called \toolname, to automatically analyze enclave binary code based on symbolic execution (see~\Cref{sec:architecture}).
We implement vulnerability detectors in \toolname\ that take all the peculiarities of SGX enclaves into account allowing developers 
to identify vulnerabilities in enclave binaries a priori, i.e., before they are utilized in production.

We especially focus our investigation on the validation of pointers that are passed from the host application to the enclave.
Our findings demonstrate that developers are not aware of the difficulties of securely implementing enclave code when dealing with the critical host-to-enclave boundary.
We found that the automatically generated checks of the Intel SGX SDK are insufficient for non-trivial pointer-based data structures and a lack of proper manual validation of pointers or pointer-heavy data structures can easily lead to memory corruption vulnerabilities.

Using \toolname, we identified several vulnerabilities in publicly available enclave binaries developed at major companies such as \intel, Baidu, and \Synaptics (see~\Cref{sec:evaluation}).
Our framework features detailed vulnerability reports significantly simplifying the construction of proof-of-concept exploits to assess the reported vulnerability.
Even if no information on the enclave is available, we are able to construct exploits (see the fingerprint enclaves analyzed in~\Cref{subsec:enclave:synatee} and~\ref{subsec:enclave:goodix}).
Our exploits hijack the enclave's control-flow, effectively bypassing all security guarantees of the SGX technology.
By performing root-cause analysis we identified five vulnerability classes that repeatedly occur in our dataset:
\vulnPtrInDS, \vulnResource, \vulnOverlap, \vulnNull, and \vulnTOCTOU.

Interestingly, among the enclaves we found vulnerable is one enclave written by Intel engineers and published as an open-source example enclave on Intel's GitHub page~\cite{intel-github-gmpdemo}.
Another interesting finding is a vulnerability in a sample SGX enclave originally developed at Baidu
with the Rust SGX SDK (now an Apache Incubator project).
Rust features memory safety and as such has the potential to eradicate memory corruption attacks.
However, the host-to-enclave boundary is inherently memory unsafe and as such, using memory-safe programming languages in SGX does not automatically result in secure enclave code.

%% file: sections/problem-sgx-memory-corruption.tex
The lack of built-in memory safety in the common system-level programming languages C/C++ has led to a multitude of memory corruption vulnerabilities in the last three decades~\cite{eternal-war}.
These vulnerabilities allow an attacker to perform a limited or (often) arbitrary write to memory.
Such malicious writes manipulate (1)~control-flow information on stack and heap (e.g., return addresses and function pointers) or (2)~so-called non-control data (e.g., decision-making variables).
In both cases, the attacker influences the program's execution flow and eventually executes a malicious sequence of instructions.
In the recent past, we witnessed an arms race between defenses and memory corruption attacks: data-execution prevention~\cite{pax-nx,DEP} effectively prevents malicious code injection in data memory, but can be bypassed by means of return-oriented programming (ROP) attacks as these only reuse code already residing in code memory~\cite{shacham2007rop}.
Software-diversity based defenses~\cite{sok-aslr,Jackson2011compilersoftwarediversity} mitigate ROP attacks by randomizing the location of code in memory but are circumvented if an attacker manages to dynamically disclose the code location~\cite{JIT-ROP}.
Similarly, control-flow integrity (CFI)~\cite{Abadi2009} depends on the precision of the control-flow graph (CFG) as CFG over-approximation opens the door for subtle ROP attacks~\cite{Goktas2014outofcontrol,Davi2014stitchingthegadgets,Carlini2014ropdangerous}.
Lastly, even if one would be able to develop a perfect CFI scheme, non-control data attacks would still be a viable attack option as they only execute execution paths that adhere to the program's CFG~\cite{Chen2005-noncontrol,Hu2016-dop,Ispoglou2018-bop,Pewny2019-steroids}.

In general, SGX enclaves are as susceptible to memory corruption attacks as any other system software.
In fact, almost all enclaves are developed in C/C++ mainly because the official Intel SGX SDK~\cite{sgxsdk-linux} provides a C/C++ development environment.
Only recently, memory-safe languages such as Rust have been explored as a programming language for SGX enclaves~\cite{Wang2019rustsgxccs}.
However, as we will show, even these cannot guarantee that enclaves are free of memory corruption vulnerabilities.

One particular challenge arises when launching memory corruption attacks against SGX enclaves: since SGX enclaves are encrypted in memory and can be shipped as an encrypted binary~\cite{schuster2015vc3,baumann2014shielding}, an attacker cannot necessarily perform static analysis on the enclave's binary to search for interesting ROP gadgets (i.e., enclave code sequences maliciously combined to trigger malicious operations).
\citewauthor{lee2017darkrop} tackle this challenge by repeatedly executing an enclave, triggering the execution at different entry points, and analyzing memory access to dynamically identify ROP gadgets.
Note that this attack does not apply to enclaves whose code addresses are randomized for each instantiation of the enclave.
On the other hand, existing SGX randomization schemes such as SGX-Shield~\cite{seo2017sgxshield} are not able to apply randomization to all of the enclave's code area: \citewauthor{Biondo2018-dilemma} demonstrated that the Intel SGX SDK provides enclave libraries that are not randomized and include several powerful ROP gadgets (i.e., gadgets that allow control of many processor registers).
Specifically, these gadgets are invoked when resuming the context of an SGX enclave (\ocall-return).
Hence, an attacker only needs to launch a memory corruption attack and  provide counterfeit context information to hijack a vulnerable enclave.

\paragraph{Problem Setting.}
We observe that existing memory corruption attacks against SGX~\cite{lee2017darkrop,Biondo2018-dilemma} exploit the host-to-enclave boundary as this serves as entry point to trigger and halt enclave execution.
Further, the existing attacks assumed that the attacker is capable of hijacking the control flow of the enclave's code by means of a given memory corruption vulnerability.
However, the open question is whether such vulnerabilities are likely to occur when developing enclaves.
To answer this question, we reverse-engineer public enclave code and develop automated analysis techniques to assess the security of enclaves regarding memory corruption vulnerabilities.
Our findings demonstrate that an erroneous implementation of the API at the host-to-enclave boundary is often the root-cause for memory corruption vulnerabilities in SGX code.

%% file: sections/background.tex
In this section, we provide background information on the Software Guard Extensions (SGX) technology of modern \intel\ processors and more specifically the \sgxsdk.
The \sgxsdk is currently the primary way to develop SGX enclave code and is officially endorsed by \intel.

\subsection{Host-Enclave Interface}

\begin{figure}[t]
	\begin{center}
		\includegraphics[width=0.8\linewidth]{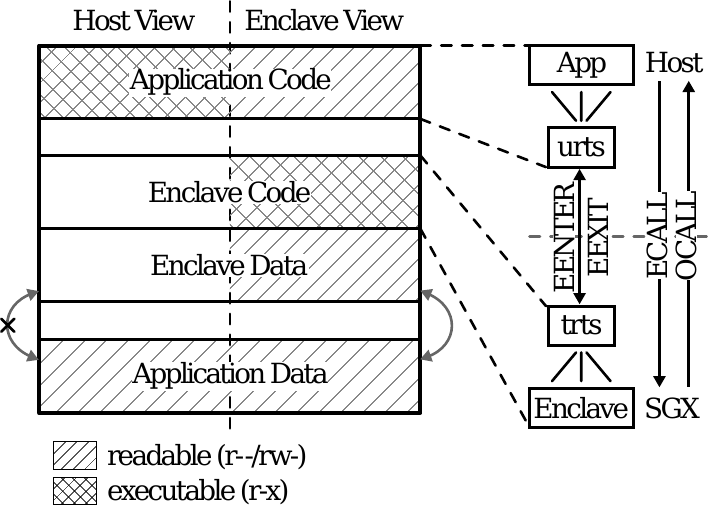}
	\end{center}
	\caption{General overview on SGX-enabled applications.}
	\label{fig:rtshw}
\end{figure}

\Cref{fig:rtshw} provides a general overview of the memory layout of SGX-enabled applications as well as the channel for host-to-enclave interaction.
The SGX enclave is part of a user space application, called host process or application, which eventually loads and executes the enclave. 
Both host and enclave share the same virtual address space with the exception that the enclave resides in encrypted and integrity-protected memory.
As shown in \Cref{fig:rtshw}, in the enclave view enclaves can access all of the host application's memory.
Only the enclave memory is assumed as trusted, whereas all other memory parts are considered as untrusted.

The host process starts the enclave's execution by issuing the special \eenter instruction to enter the enclave.
For this, enclaves define entry points in the so-called \emph{thread control structures} (TCS), which are locked while in use by a thread.
This makes the number of TCS also the maximum number of threads that can enter an enclave concurrently.
A jump from the executing enclave to code in the host application results in a segmentation fault, effectively making host code non-executable for the enclave.
As such, the enclave must explicitly leave enclave mode by using the \eexit instruction before the thread can execute any non-enclave code.

The \sgxsdk\ provides the concept of \emph{\ecalls} (enclave calls) on-top of \eenter to control the transition from application code to enclave code.
For a simple enclave, not requiring multi-threading, the SDK uses only one TCS which is called with the index of the desired \ecall.
First, the host application calls the \ecall wrapper in the untrusted runtime (urts). Next, the urts prepares the transition to the SGX enclave according to the so-called EDL file.
Second, it executes the \eenter instruction to transfer control to the enclave code.
More specifically, control is transferred to an enclave entry point in the SDK's trusted runtime (trts).
The trts takes care of the context switch and sets up the enclave execution environment:
\begin{inparaenum}[(1)]
\item it switches the stack to a stack in enclave memory,
\item allocates secure memory and copies the arguments into the enclave,
\item calls the actual \ecall function, and finally
\item clears the registers before returning to the host application's code.
\end{inparaenum}
Similarly, the SDK also supports calling functions of the untrusted host application, which is referred to as \emph{\ocalls} (outside calls).
For \ocalls, the trts saves the enclave's execution state to enclave memory and restores it when the call returns.

\subsection{The EDL Interface Specifications}
\label{sec:edl}

The \sgxsdk\ uses the EDL (\emph{Enclave Definition Language}), a custom specification language to define the \ecall and \ocall interface of an enclave.
The EDL language resembles a C-header file with additional syntax to specify SGX-specific information.
It allows the developer to specify the prototypes of functions available as \ecall and the valid data format of input arguments.
Based on the EDL file, the SDK generates wrapper code to transparently connect the function stubs in the host application with the \ecalls in the enclave.
The parameters of \ecalls are transferred using auto-generated data structures.
When the application invokes an \ecall, the SDK-generated code stores all parameters in the prepared structure in the untrusted host application memory.
These will then be fetched by the SDK code in the enclave and unpacked for the actual \ecall code.

The SDK must be able to determine the size of the arguments to allocate a fitting buffer in the secure memory.
Thus, every pointer type has to be annotated with a size such that the SDK can determine the size of the underlying buffer.
Currently, the \sgxsdk\ supports copying C data types such as basic integer types, composed basic data types (\emph{struct}) without nested pointers, 0-terminated/C-style strings, and pointers to arrays of fixed length.

\begin{figure}[t]
\centering
\begin{lstlisting}[language=EDL]
enclave {
  trusted { // ECALLs
    public void ecall_size1( // explicit size
        [in, size=100] void* ptr);
    public void ecall_size2( // variable size in len
        [in, size=len] void* ptr, size_t len);
    public void ecall_user( // dangerous user_check
        [user_check] void* ptr);
  };
};
\end{lstlisting}
\caption{Example for the EDL syntax.}
\label{fig:edl-example}
\end{figure}

\Cref{fig:edl-example} shows an example of different features of the EDL language.
In this example, a \verb+void*+ pointer is annotated with \verb+[in, size=100]+.
The SDK will generate code that allocates 100\,bytes in enclave memory and copies 100\,bytes from untrusted memory into the enclave.
Alternatively, the developer can also specify dynamic lengths, which then refer to other parameters by name.
However, when writing the interface definition in EDL, there are some peculiarities that have to be taken into account.
First, it is possible to disable the SDK features. 
A pointer that is annotated as \sym{[user\_check]} is passed to the enclave without any auto-generated check. 
It is up to the enclave developer to validate the underlying buffer.
Second, compound data types are only shallow copied. 
They are treated as buffers with a fixed size and are simply copied into secure memory. 
Data structures are not recursively copied, i.e.\ it is not checked if any of the fields in the structure is a pointer type.
So, even if a developer uses the \sgxsdk to protect the \ecall API, there are many cases that additionally require custom validation code, which is error-prone.

%% file: sections/symex-implementation.tex
\begin{figure*}[ht!]
	\begin{center}
		\includegraphics[width=0.8\textwidth]{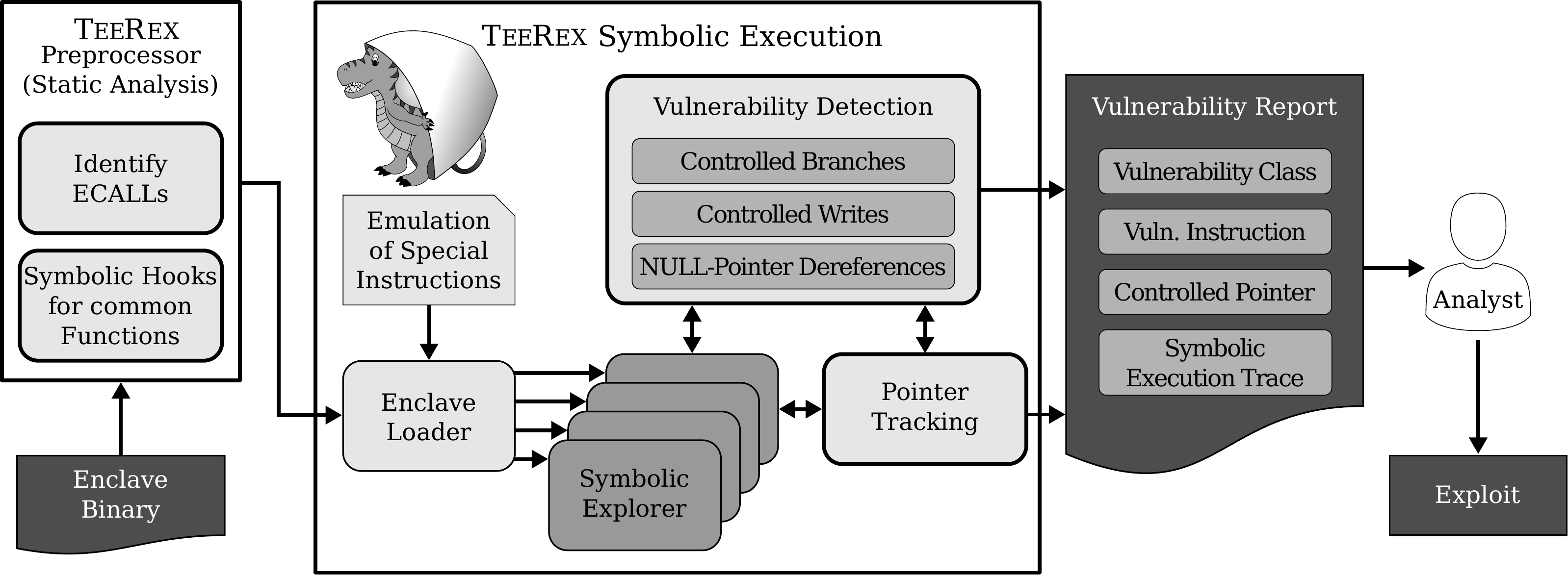}
	\end{center}
	\caption{Architecture of \toolname}
	\label{fig:architecture}
\end{figure*}

We develop a novel symbolic execution framework, called \toolname\nowidth{,}\footnote{\toolname stands for \emph{\longtoolname}} to \emph{automatically} identify vulnerabilities of SGX enclaves.
Our framework does not only identify vulnerabilities, but also generates a detailed vulnerability report which significantly simplifies the process of constructing proof-of-concept exploits against the vulnerable SGX enclave.
It supports all platforms supported by the \sgxsdk: Windows~(PE) and Linux~(ELF) binaries and both 32 and 64-bit enclaves.
Note that we apply symbolic execution on the \emph{binary level} to be able to analyze closed-source, proprietary enclaves.
Our prototype of \toolname\ supports the standard enclave format of the \sgxsdk and leaves support of custom enclave formats and loaders (e.g., the Graphene framework~\cite{graphene}) as future work.
Further, we focus our analysis on unencrypted enclave code.
In case the enclave code is encrypted neither \toolname nor any other static analysis tool can analyze the enclave without knowing the secret key.
\toolname must be able to read and properly load the enclave's code.

In what follows, we describe the overall architecture of \toolname (Section~\ref{sec:impl-symex}), elaborate on several challenges and how we tackled them (Section~\ref{sec:impl-challenges}), and finally describe our vulnerability detection engines in detail (Section~\ref{sec:impl-engines}).
\subsection{Architecture}%
\label{sec:impl-symex}

Symbolic execution was first proposed in the 70s as a generalization of testing~\cite{King1976symex,Boyer1975selectsymex} and has become one of the standard tools for high coverage testing and vulnerability analysis~\cite{Baldoni2018symexsurvey,shoshitaishvili2016state,Cadar2008klee,Cha2012mayhem}.
However, the modeling of side effects caused by the operating system (OS) is highly challenging, e.g., symbolic execution must typically simulate and support all OS system calls and manage a simulated file system~\cite{Baldoni2018symexsurvey}.
Fortunately, there are several SGX peculiarities that simplify symbolic execution for SGX enclaves: enclave code is self-contained (i.e., no external dependencies like libraries) and isolated from the rest of the system.
SGX enclaves are prohibited to perform any system calls and any interaction with the OS is handled by means of an \emph{\ocall} to the untrusted host application.

Figure~\ref{fig:architecture} shows the architecture of \toolname' symbolic analysis pipeline.
The main goal of \toolname is to find vulnerable states during the symbolic exploration. Further, it aims to collect meta-data to eventually generate a detailed vulnerability report.
This is achieved by executing each \ecall symbolically and checking every state for different vulnerability classes. %
To produce accurate vulnerability reports, we add pointer tracking to the symbolic execution engine. This allows us to track pointer dereferences and propagate labels that allow us to distinguish between data loaded from enclave and host memory.
As a result, \toolname can spot vulnerable instructions that read data from outside of the protected enclave memory.
This is a necessary design decision as enclaves can be loaded by arbitrary (malicious) host applications.

We leverage the well-known \angr~framework~\cite{shoshitaishvili2016state} as our symbolic explorer. This allows us to extract memory constraints from enclave code, which is subsequently needed for vulnerability analysis.
\angr itself does not support executing SGX enclaves because:
\begin{inparaenum}[(1)]
	\item \angr cannot jump from the host application to the enclave
	\item there is no setup for an initial environment to directly execute \ecalls,
	\item enclaves utilize CPU instructions not supported by \angr,
	\item \toolname leverages enclave specifics to scale over multiple processes and machines, while \angr is limited to one thread, and
	\item the common trusted functions for memory allocation are not directly supported by \angr.
\end{inparaenum}
Furthermore, \angr does not perform any vulnerability analysis by itself: its purpose is to provide a robust and comprehensive framework to perform static analysis and symbolic execution.
As we will describe in Section~\ref{sec:impl-challenges}, \toolname tackles all the above mentioned challenges.
As shown in \Cref{fig:architecture}, \toolname is split into several major components.

\component{Preprocessor}
The first step in the pipeline depicted in \Cref{fig:architecture} is to pre-process the enclave binary to
\begin{inparaenum}[(1)]
	\item identify instructions and functions that cannot be executed symbolically, and
	\item to locate the \ecall table and extract the addresses of the \ecall functions.
\end{inparaenum}
This preliminary static analysis step allows us to instrument specific binary instructions to increase the performance and coverage of the analysis.

\component{Enclave Loader}
The enclave loader sets up the initial environment to execute one \ecall. It replaces the identified \emph{common functions} and \emph{special instructions} with emulating Python code. Further, it creates the argument structure for the \ecall with unconstrained %
symbolic values.

\component{Symbolic Explorer}
The symbolic execution performed by the \angr~framework can be distributed across multiple machines, as the \ecalls are analyzed individually. The results are merged later in the vulnerability reports for the analyst.

\component{Vulnerability Detection}
\toolname analyzes the symbolic states during \angr's symbolic exploration for vulnerabilities in the enclaves.
It specifically analyzes instructions that access memory and jumps.
This is described in detail in \Cref{sec:impl-engines}.

\component{Pointer Tracking}
The majority of vulnerabilities in SGX enclaves are due to insecure pointer usage and lack of pointer validation.
\toolname implements pointer tracking by analyzing all pointer dereferences and propagating labels between symbolic values.
More specifically, \toolname uses a taint-style analysis annotating every value loaded from memory with the address, where the value was loaded from.
This allows \toolname to determine the source of a value, e.g., whether a function pointer used for an indirect call was loaded from enclave, host memory, or loaded via a parameter passed to the \ecall function.

Furthermore, \toolname places hooks on \sgxsdk functions that are used to validate whether an address is within secure memory.
Whenever the enclave uses one of these functions, \toolname forks the symbolic execution into two states: one where the address is within enclave memory and one where the address is outside enclave memory.
This information is used by \toolname to assess whether a bug is exploitable and report identified vulnerabilities more accurately.

\component{Vulnerability Report}
Finally, \toolname\ produces a vulnerability report, which contains
\begin{inparaenum}[(1)]
	\item the type of the vulnerability,
	\item the location in the binary,
	\item the controlled pointer and its position in the attacker-controlled input and
	\item an execution trace to reach the vulnerable instruction.
\end{inparaenum}
The vulnerability report provides sufficient detailed information to an analyst for constructing a proof-of-concept exploit, even for closed-source enclaves (see Section~\ref{sec:evaluation}).

\subsection{Challenges}%
\label{sec:impl-challenges}

Next, we will describe several challenges when applying symbolic execution to enclave binaries and how our design tackles them.

\challenge{Accuracy and Scalability}
Enclaves built with the \sgxsdk define only a few (often one) entry point in the thread-control structure (TCS).
This entry point is the trusted runtime (trts) that is responsible for setting up the enclave execution environment, calling exception handlers, and multiplexing \ecalls.
For this, the arguments of an \ecall are packed by the untrusted runtime (urts) to be unpacked upon entering the enclave by the trts.
This introduces an additional layer of pointer indirection for all \ecall parameters.
The specifics of the enclave management in the trts are heavily dependent on the intrinsics of the SGX instructions and the enclave's internal metadata, which are not present in the emulated environment. This introduces high complexity and a major challenge for a symbolic execution analysis because
\begin{inparaenum}[(1)]
	\item the enclave initialization routines result in many memory accesses through symbolic addresses, which is a notoriously hard problem for symbolic execution engines in general~\cite{Baldoni2018symexsurvey,Cha2012mayhem}, and
	\item due to the low-level nature of the trts code the symbolic execution lacks semantic information about the execution context when it finally reaches the \ecall functions.
	Hence, it is not feasible to map symbolic memory ranges to \ecall parameters once the symbolic execution analyzes the actual \ecall function.
\end{inparaenum}

However, symbolically executing the whole trts code is conceptually uninteresting for identifying vulnerabilities in \ecalls as the trts is independent of \ecalls.
As such, we designed \toolname in such a way that it is able to skip symbolic execution of the trts and instead targets \ecall functions directly.
To do so, \toolname first extracts the \ecall table from the enclave binary.
Next, symbolic execution is started at the beginning of every \ecall separately.
This allows \toolname to produce very accurate vulnerability reports as it is now possible to directly control the arguments passed to the \ecall function. At the same time, it reduces the overhead of executing code that is not meaningful for identifying exploitable bugs in enclaves.
Furthermore, starting the analysis for each \ecall function separately and skipping the SDK runtime components allows parallelization of the symbolic execution process.
Note that \angr is originally restricted to one thread due to the limits of the Python implementation.

\challenge{Standard Memory Functions}
Another source of path complexity arises from the standard memory functions.
Methods like \sym{memcpy} or \sym{malloc} are reimplemented in \angr\ as so-called \emph{SimProcedures} at a higher level.
Instead of symbolically executing the binary code of a function like \sym{memcpy}, \angr\ instead invokes the corresponding SimProcedure to update the symbolic state.
This is possible because most applications load these functions dynamically from a library in the system, which can be easily intercepted.
However, the self-contained enclave code comes with its own trusted version of these functions.
As such, \toolname searches the enclave code for trusted versions of these functions and places hooks to invoke the corresponding SimProcedure instead.

\challenge{Unsupported CPU Instructions}
Since SGX has been recently integrated into new Intel CPUs, there are several advanced instructions included in enclave code that are unsupported by the symbolic explorer either because they are too new or too complex to be easily implemented symbolically.
This includes the primary SGX instruction \sym{enclu} to enter/exit an enclave, but also the non-SGX-specific instructions \sym{rdrand} and \sym{xsave/xrstor}, which are used in \ocalls to save and restore all registers from memory when the execution passes the host-enclave boundary.
To tackle this challenge, we avoid executing the SGX-specific entry instructions but directly invoke the \ecall functions during the symbolic execution.
We deal with other unsupported, but frequently executed instructions, by hooking into them.
The hooks re-implement and emulate the instructions in Python to update the symbolic state accordingly.

\challenge{Global State of Enclaves and Chains of \ecalls}
Enclaves can be entered multiple times at different \ecalls with different attacker-controlled input data, with each of the calls altering the internal global state of the enclave.
Hence, the control-flow of an \ecall does not only depend on its arguments, but also on all prior invoked \ecalls.
Taking this into account, an accurate symbolic exploration of an \ecall requires exhaustive knowledge about the effects of all \ecalls.
To address this issue, \toolname analyzes each \ecall individually and treats all (secure) global state (i.e., global variables in the \sym{data} and \sym{bss} sections) of an enclave as initialized with unconstrained symbolic values.
This allows our tool to also explore paths of an \ecall that are not reachable with an enclave's initial global state.
However, the global state is typically not fully attacker-controlled but rather initialized to zero or changed to some value by a different \ecall.
Thus, the assumption that the global state is completely unconstrained can potentially lead to a situation, where our \toolname\ wrongly reports an attacker-controlled jump or write although the state might be limited to only safe values.
Nevertheless, the analysis results are still useful because they can lift limited exploitation primitives (e.g., null-pointer dereference or write to an arbitrary address with a fixed value) to full control-flow hijacking attacks (see \Cref{subsec:enclave:synatee} for an example).

\subsection{Vulnerability Detection Components}%
\label{sec:impl-engines}

We implemented three major vulnerability detection components in \toolname:
\begin{inparaenum}[(1)]
  \item attacker-controlled branches (control-flow hijacking),
  \item controlled writes, and
  \item NULL-pointer dereferences.
\end{inparaenum}
To analyze an enclave, \toolname\ first reads the \ecall table from an enclave and symbolically executes the \ecall functions sequentially.
We pass fully symbolic arguments to each \ecall function and symbolically explore its code.
Our symbolic execution tool currently supports detecting two major classes of exploit primitives: control-flow hijacking and controlled writes.
In addition, we detect if the enclave dereferences a NULL-pointer.

\paragraph{Control-Flow Hijacking.}%
To identify control-flow hijacking vulnerabilities, \toolname searches for program paths, where the enclave utilizes attacker\hyp{}controlled data as a jump target.
To be more precise, \toolname\ detects and reports unconstrained jumps that are encountered during symbolic execution.

Anything that is attacker\hyp{}controlled (i.e., input and the whole address-space outside of enclave memory) is marked as an unconstrained fully symbolic value during symbolic execution.
This means that when the \ecall uses one of its symbolic arguments as a jump target, it will jump to an unconstrained symbolic value.
Furthermore, loading the jump target from uninitialized memory also leads to loading an unconstrained symbolic value.
On the other hand, if the enclave validates the jump target pointer to be within a certain set of allowed values, then the symbolic execution engine will gather constraints on the symbol representing the jump target during the analysis of the validation code.
The jump target is now tightly constrained to be within a certain set of allowed---assumed to be safe---values, which will not trigger an alarm.
However, any use of an unconstrained pointer as a jump target results in \toolname reporting a controlled jump, as here no prior validation was found and the attacker has full control over the jump target.

\paragraph{Controlled Write.}
\toolname searches for writes to arbitrary (unconstrained) memory addresses during symbolic exploration.
Therefore, we track every pointer dereference and propagate labels similar to taint analysis~\cite{clause2007dytan,vanderveen2017newton}.
This makes it possible to infer the relation of a corrupting pointer to the input arguments. 
This includes the levels of indirection and corresponding offsets.
When a pointer is utilized for a memory write, \toolname\ checks whether the address is related to attacker-controlled memory.
If the address was loaded based on input arguments, the attacker can directly control the address used in the memory write instruction.
Furthermore, \toolname uses the solver of the symbolic execution engine to test whether the address of a write can possibly point to an arbitrary memory location inside of the enclave memory.
If so, we can infer that we discovered an arbitrary write gadget.

Any write to an arbitrary address must be considered as a vulnerability regardless of whether the value written is attacker\hyp{}controllable.
For instance, a controlled write to an arbitrary address with a fixed single byte value (e.g., \hex{0x0a}) is often sufficient to corrupt a pointer in enclave memory.
With complete control of the address space in the SGX setting, the attacker can map memory pages at almost any address.
As a result, it is sufficient if the attacker can partially corrupt a pointer in enclave memory and make it point to insecure memory, which still is a valid memory location (see exploit in \Cref{subsec:enclave:synatee} for an example).
As such, \toolname reports any memory write to an attacker controlled address, regardless of the value written.

\paragraph{NULL-Pointer Dereference.}
On the x86~architecture, the page at address 0~(NULL) in the virtual address space of a user space program is a legitimate address.
However, in C/C++, pointers are typically initialized to the null-pointer and many functions from standard libraries return the null-pointer to indicate an error.
As such, dereferencing a null-pointer is a common problem in C/C++ code but typically not considered critical as the null page is not mapped, i.e.\ the process only crashes when trying to dereference a null-pointer.
On the contrary, in the SGX setting a null-pointer dereference is critical since the null page is typically not within trusted enclave memory.
As such, we need to consider it as controlled by the attacker.
\toolname\ analyzes every memory access and checks whether the address is pointing to the zero page mapped at address~$0$ (typically $<$~\hex{0x1000}).
If this is the case, \toolname reports that the code is dereferencing a null-pointer.

%% file: sections/results.tex
\input{tables/enclaveoverview.tex}

To evaluate the effectiveness of \toolname\ on real-world enclaves, we gathered a dataset consisting of open-source and proprietary public enclaves.
Table~\ref{tab:enclaveversions} provides an overview of all the enclaves we analyzed with \toolname. 
Our dataset contains enclaves developed by well-known companies such as Intel and Baidu.
We also included SGX-protected fingerprint software that is utilized in \Dell and \Lenovo laptops. 
Note that it was highly challenging finding projects utilizing the SGX technology.
We assume this is due to the fact that SGX is a rather new technology, hardware-support on client machines is still not widely available, and as such, SGX is primarily used in cloud settings where the enclave is simply not publicly available.

We use the following methodology for analyzing the enclaves in our dataset:
first, we analyze the enclaves with \toolname.
Second, using the vulnerability report of \toolname, we verify the vulnerabilities, perform root-cause analysis to identify the vulnerability, and finally construct a proof-of-concept (PoC) exploit.
In our PoC exploits, we aim to hijack the instruction pointer while the processor is in enclave mode.
Given this capability, an attacker can utilize existing code-reuse attack techniques to achieve arbitrary code execution~\cite{Biondo2018-dilemma,shacham2007rop}.
By constructing such a PoC exploit, we gain confidence that the issues discovered by \toolname are indeed serious vulnerabilities.

For our PoCs we assume the standard SGX adversary model~\cite{Costan2016-sgxexplained,McKeen2013-hasp-sgx-apps} in which the attacker has full control over the user space and operating system/hypervisor.
More specifically, our current PoCs assume a standard OS (Ubuntu 18.04 and Windows 10), which are configured or patched to allow the attacker to map the page at address~$0$. 
The enclaves are all compiled with the standard \sgxsdk.
Note that our PoCs do not need to bypass ASLR since the untrusted OS selects the address space layout of the enclave.
Our PoC exploits attempt to get full control over the instruction pointer, which is typically sufficient to perform arbitrary code execution~\cite{Biondo2018-dilemma}.

Using \toolname, we identified vulnerabilities in all of our analyzed SGX enclaves except the SignalApp contact discovery service~\cite{signal-enclave}. 
In our analysis, we observed that the \ecall interface of this enclave is comparatively small and simple.
For each of the vulnerable enclaves, we successfully developed PoC exploits of which all enable full instruction pointer control.
We performed responsible disclosure for all vulnerable enclaves listed in Table~\ref{tab:enclaveversions}. 
All vendors have acknowledged our findings and all vendors, except for one, developed fixes for the vulnerabilities we reported.

We also performed root-cause analysis on our findings and identified several problematic code patterns that lead to vulnerabilities.
Table~\ref{tab:enclavevulns} shows an overview of the results of our analysis.
We identified and successfully abused all different types of exploit primitives that \toolname detected.
Based on our root-cause analysis we identified bug classes specific to SGX that easily lead to vulnerabilities in enclave code.
In what follows, we discuss in detail the vulnerable enclaves and bug classes we identified.

\begin{table*}[ht]

\input{tables/enclavevulns.tex}
\label{tab:enclavevulns}
\end{table*}

\subsection{\enclaveIntelGMP}%
\label{subsec:enclave:intel}

\input{sections/enclaves/intel.tex}
\subsection{\enclaveWolfSSL}%
\label{subsec:enclave:wolfssl}

\input{sections/enclaves/wolfssl.tex}

\subsection{\enclaveRustTLS{}/server}%
\label{subsec:enclave:rust}

\input{sections/enclaves/rust.tex}

\subsection{\enclaveTalos}%
\label{subsec:enclave:talos}

\input{sections/enclaves/talos.tex}

\subsection{\enclaveSynaTEE}%
\label{subsec:enclave:synatee}

\input{sections/enclaves/synatee.tex}
\subsection{\enclaveGoodix}%
\label{subsec:enclave:goodix}

\input{sections/enclaves/goodix.tex}

\subsection{Vulnerability Disclosure}%
\label{subsec:disclosure}

\input{sections/disclosure.tex}

%% file: tables/enclaveoverview.tex
\begin{table*}[ht]

\newcommand{\tyes}{$\checkmark$}
\newcommand{\tno}{$\times$}
\newcommand{\tna}{n/a}

\newcommand{\lb}[2]{\parbox{\widthof{#1}}{\centering #2}}

\begin{center}
\begin{tabularx}{\linewidth}{X l c l c l c}
  \toprule
  Project Name & \lb{Analyzed}{Analyzed Version} & Exploit & Fixed Version(s) & \lb{Source}{Source Code} & Target & \lb{Number of}{Number of \ecalls} \\
  \midrule
  \enclaveIntelGMP~\cite{intel-github-gmpdemo} &  \gitcommit{9533574f95b97ea08adb6724d8be797c53119dac} & \tyes & \gitcommit{0491317b4112b06e16b3f3b1c07b06e400b32391} & \tyes & Linux amd64 & 6 \\
  \enclaveRustTLS~\cite{rust-sgx-tlsclient,Wang2019rustsgxccs} & \version{1.0.9} & \tyes & \gitcommit{f975a19982740d5d2e878b595c1be5d1a1a31ecb} & \tyes & Linux amd64 & 8 \\
  \enclaveTalos~\cite{talos,talos-report} & \gitcommit{bb0b61925347b5148fe44cd6400eb981bd0f5a36} & \tyes & not planned & \tyes & Linux amd64 & 207 \\
  \enclaveWolfSSL~\cite{wolfssl-example-enclave} &  \gitcommit{d330c53baff52fdf4338619cd4f82ae25c1bc294} & \tyes & \gitcommit{1862c108d7e3be47a3d6fe18f406df444ae36e6e} & \tyes & Linux amd64 & 22 \\
  \enclaveSynaTEE & \enclaveSynaTEEversion & \tyes & {\cite{synaptics-cve,synaptics-advisory,lenovo-advisory,hp-advisory}} & \tno & Windows amd64 & 2 (76)\textsuperscript{*} \\
  \enclaveGoodix & \enclaveGoodixVersion & \tyes & {\cite{goodix-cve,dell-advisory}} & \tno & Windows amd64 & 56 \\
  \enclaveSignal~\cite{signal-enclave} & \version{1.13} & \tno & - & \tyes & Linux amd64 & 7 \\
  \bottomrule
\end{tabularx}
\end{center}

\let\tyes\undefined
\let\tno\undefined
\let\tna\undefined

\vspace{-1em}
\caption{Dataset of public enclaves and their susceptibility to exploitation.}%
\label{tab:enclaveversions}
\centering\footnotesize
\textsuperscript{*}~One \ecall immediately branches to 75 different actions.
\end{table*}

%% file: tables/enclavevulns.tex
\newcommand{\tyes}{$\bullet$}
\newcommand{\tno}{-}
\newcommand{\rota}[1]{\rotatebox[origin=l]{90}{#1}}
\newcommand{\maxw}[1]{\parbox{2.8cm}{\raggedright{}#1}}
\newcommand{\head}[1]{\rota{\maxw{#1}}}

\newcommand\ta{}
\newcommand\tb{}

\renewcommand{\head}[1]{\makebox[2em][l]{\rotatebox[origin=l]{30}{#1}}}

\begin{center}
\begin{tabular}{ c | r | c c c c c c l }
  \toprule
  & & \head{\enclaveIntelGMP} & \head{\enclaveRustTLS} & \head{\enclaveTalos} & \head{\enclaveWolfSSL} & \head{\enclaveSynaTEE} & \head{\enclaveGoodix} & \hspace{6em} \\
  \midrule
  \multirow{5}{*}{\rota{Bug Classes}}%
  & \titlevulnPtrInDS     & \tyes   & \tyes & \tyes   & \tno  & \tyes   & \tyes   \\
  & \titlevulnResource    & \tyes   & \tyes & \tyes   & \tyes & \tno\ta & \tno\ta \\
  & \titlevulnOverlap     & \tno\tb & \tyes & \tno\tb & \tno  & \tno\tb & \tno\tb \\
  & \titlevulnNull        & \tno    & \tno  & \tyes   & \tno  & \tyes   & \tyes   \\
  & \titlevulnTOCTOU      & \tno    & \tno  & \tyes   & \tno  & \tno    & \tno    \\
  \midrule
  \multirow{3}{*}{\rota{\parbox{\widthof{Primitive}}{\centering{}Exploit\\Primitive}}}%
  & Control-Flow Hijack      & \tno  & \tyes & \tyes & \tyes & \tyes & \tyes \\
  & Controlled Write         & \tyes & \tno  & \tno  & \tno  & \tyes & \tyes \\
  & NULL-pointer Dereference & \tno  & \tno  & \tyes  & \tno  & \tyes & \tyes \\
  \bottomrule
\end{tabular}
\end{center}
\vspace{-1em}
\caption{%
Overview of results of our analysis of public enclave code.%
\ Some patterns are not applicable for every enclave, because the relevant code constructs are not used or the source is unavailable.
}%

\let\tyes\undefined
\let\tno\undefined
\let\rota\undefined
\let\maxw\undefined
\let\head\undefined
\let\a\undefined

%% file: sections/enclaves/intel.tex
Intel provides the \emph{GNU Multiple Precision Arithmetic Library} for SGX and a corresponding demo application.
The enclave code takes two GMP big integers as parameters, performs an arithmetic computation, and returns the result.
\toolname\ identified an arbitrary write vulnerability in the enclave code, which we used in our PoC exploit to gain arbitrary code execution.
The data structure behind the GMP big integer internally utilizes a pointer to refer to an underlying buffer that contains the variably-sized data of the big integer.
\toolname\ identified that this pointer is not sanitized allowing a memory write to an arbitrary location.
This vulnerability shows how likely it is for SGX developers utilizing a third-party library, to miss validating a pointer inside of opaque data structures.

\begin{figure}[t] \centering \begin{lstlisting}
void e_mpz_add(mpz_t *c_unsafe,
               mpz_t *a_unsafe,
               mpz_t *b_unsafe) {
  mpz_t a, b, c;
  /* [computation code omitted] */
  // mpz_set copies the underlying buffer
  // of the biginteger "c" to the buffer pointer
  // contained in the "c_unsafe" variable
  mpz_set(*c_unsafe, c);
}
\end{lstlisting}
\caption{Excerpt of the vulnerable code in the \emph{\enclaveIntelGMP} enclave.}
\label{fig:gmp-vuln-code}
\end{figure}

The problem behind the vulnerability is that the numbers passed to the enclave are GMP big integer objects representing arbitrary large integers.
The GMP big integer data structures utilize dynamically allocated storage internally; they contain a pointer to the underlying buffer that stores the actual integer value.
However, the enclave fails to properly validate the pointer inside of the GMP data structure.
Figure~\ref{fig:gmp-vuln-code} shows part of the vulnerable code:
the enclave receives three big integer parameters.
The first one, called \sym{c\_unsafe}, is used as an output parameter.
The enclave uses functionality of the GMP library that is not SGX-aware: the \sym{mpz\_set} function.
As such, the library function simply copies the output to the attacker-controlled underlying buffer of the \sym{c\_unsafe} big integer.
This neglects the fact that the underlying buffer of this big integer can actually point to arbitrary memory, including enclave memory.

This vulnerability allows an attacker to perform an arbitrary memory write, with controlled content and controlled size.
\toolname identifies the arbitrary write vulnerability in multiple \ecalls. They all share the same structure as the one depicted in Figure~\ref{fig:gmp-vuln-code}.
In our proof-of-concept exploit, we abuse the \sym{e\_mpz\_add} \ecall:
we set the value of the underlying buffer of the big integer parameter \sym{a\_unsafe} to our payload, the big integer \sym{b\_unsafe} to a big integer initialized as~0, and the underlying buffer of \sym{c\_unsafe} to our target address for the arbitrary write.
We choose an address on the enclave stack that points to a return address used by the enclave.
This effectively allows us to write a ROP-payload directly onto the enclave stack.

Intel acknowledged the problem, updated their documentation, and fixed the issue by using serialization:
instead of passing pointers to GMP structures, the demo code now serializes GMP big integer objects to strings and passes those strings over the host-to-enclave boundary.
The enclave then deserializes the data structure, computes the result, and finally returns the serialized result back to the host application.
Since no longer GMP big integer pointers are passed between the host and enclave, this fixes the vulnerability and removes the problematic pattern \vulnPtrInDS from the enclave code, which is defined in the following:

\pattern{\titlevulnPtrInDS}
This type of vulnerability occurs due to complex data types in C/C++ that are using pointers as their primary mechanism to form complex data structures like lists, trees, or maps.
When programming with the \sgxsdk, the interface provided by an enclave allows utilization of complex data types using pointers.
However, currently the \sgxsdk does not automatically perform a recursive copy/validation of pointer-heavy data structures.
As a consequence, it becomes dangerous to pass data structures containing pointers to an enclave.
Any data structure containing pointers must be treated the same way as pointers annotated with the \verb+[user_check]+ attribute.

%% file: sections/enclaves/wolfssl.tex
WolfSSL~\cite{wolfssl} is a small TLS/SSL library without external dependencies designed for embedded devices and applications that require to be small and self-contained. It also features SGX support.
The wolfSSL project offers an enclave that showcases how to use the wolfSSL TLS library within SGX.
The enclave allows the host application to terminate a TLS connection within the SGX enclave thereby protecting all cryptographic secrets used by TLS.
However, the enclave exposes a large subset of the WolfSSL API via the \ecall interface.
We analyzed the enclave with \toolname\ and discovered a control-flow hijacking primitive in the enclave.
Our root-cause analysis revealed the following pattern, which is common to all the TLS enclaves we analyzed.

\pattern{\titlevulnResource} 
We observed that many enclaves provide functionality to allocate a resource in secure memory, e.g., a TLS session or a file object, and then return a reference to this resource to the host application.
The next time the host application attempts to use this resource, the corresponding function of the enclave is called with that reference as a parameter.
In C/C++ code, this is typically achieved by returning and passing a pointer to the object containing the resource's data.
The enclave typically validates that the given pointer indeed points to secure memory.

In the case of wolfSSL, the legacy API of the TLS library was almost directly accessible through the \ecall API of the \enclaveWolfSSL, only secured by the in-secure\hyp{}memory check, which still entailed passing the pointers of the TLS context, TLS session, and I/O buffer objects between host and enclave.
These data structures are part of a legacy API which were not designed with a split trust model in mind and it is very hard for the enclave to thoroughly validate the pointers forwarded to the legacy interface.
Figure~\ref{fig:wolfssl-code} shows the definition of the \ecall interface: a pointer to a \sym{WOFLSSL} structure is passed with the \verb+[user_check]+ attribute.
Note, that the \sym{WOLFSSL} data structure contains a function pointer used for issuing callbacks in the TLS library (\sym{CBIOSend}).
\toolname identified a control-flow hijacking primitive by passing a fake \sym{WOLFSSL} data structure with an attacker-controlled \sym{CBIOSend} function pointer.

\begin{figure}[t]
\centering
\begin{lstlisting}[mathescape,language=EDL]
/* ECALL Definition in EDL */
// a pointer to enclave memory returned
public WOLFSSL* enc_wolfSSL_new([user_check] WOLFSSL_CTX* ctx);
// pointer is passed to enclave
public int enc_wolfSSL_connect([user_check]WOLFSSL* ssl);
// ...
\end{lstlisting}
\begin{lstlisting}[language=C]
/* C Source Code */
typedef int (*CallbackIOSend)(WOLFSSL *ssl, char *buf, 
                              int sz, void *ctx);
/* WolfSSL session type */
struct WOLFSSL {
    WOLFSSL_CTX*    ctx;
    /* ... */
    // attacker-controlled function pointer!
    CallbackIOSend  CBIOSend;
}
// ...
int enc_wolfSSL_connect(WOLFSSL* ssl) {
 @\circleone{}\,@ if(sgx_is_within_enclave(ssl, wolfSSL_GetObjectSize()) != 1)
        abort();
    /* ... */ }
\end{lstlisting}
\caption{Relevant parts of the EDL definition and C source code of the \emph{tlsclient} enclave. Note the insufficient validation~\circleone.}
\label{fig:wolfssl-code}
\end{figure}

However, the \enclaveWolfSSL still implements a pointer validation routine: it validates that the pointer does point to enclave memory (\Cref{fig:wolfssl-code}:~\circleone).
However, this pointer validation is not sufficient to protect the enclave.
It is common that an attacker can actually control parts of the enclave memory, simply by providing input arguments.
For example, an attacker can abuse a different \ecall with a buffer parameter to force the enclave to copy arbitrary data into enclave memory.
In our PoC exploit, we abused the function \sym{enc\_wolfSSL\_CTX\_use\_PrivateKey\_buffer} to copy a fake \sym{WOLFSSL} structure into unrelated enclave memory (a simple buffer).
Thereafter, we call the function \sym{enc\_wolfSSL\_connect}, which uses the attacker-controlled \sym{CBIOSend} function pointer in the fake data structure, which now resides in secure memory.

This could either be fixed by using session identifiers as it was done by the \enclaveRustTLS enclave~(cf.~\Cref{subsec:enclave:rust}) or---to not change the external API---by saving all created session pointers in secure memory and only accepting these known pointers.

%% file: sections/enclaves/rust.tex
The Rust SGX SDK~\cite{rust-sgx-tlsclient} aims at introducing memory safety for SGX. As such, enclaves developed with this framework should very unlikely suffer from memory corruption bugs.
To validate this, we analyze code shipped with the Rust SGX SDK that shows how to run a TLS server and client inside of an SGX enclave.
The code consists of two similarly structured applications and enclaves that interconnect using TLS to send an HTTP request.
This shows how secure communication can be achieved while secret keys remain in protected memory.
Since both applications are similar in terms of their enclave interfaces, we only discuss the \emph{tlsclient} enclave.
The enclave API consists of functions to create a new TLS session and then utilize the session to send and receive data securely.
\toolname discovered a control-flow hijacking primitive in the enclave function \sym{tls\_client\_write} that abuses the session pointer parameter of the \ecall.
The root cause for the vulnerability of this enclave is the same pattern that already made the \nameref{subsec:enclave:wolfssl}~(\Cref{subsec:enclave:wolfssl}) vulnerable (\vulnResource).
The TLS session object is allocated within \emph{enclave memory} with the \sym{tls\_client\_new} function and then passed to further API calls like \sym{tls\_client\_write}.
The pointer has to be marked as \sym{user\_check}. Otherwise, the SGX SDK would reject the raw pointer.
However, there is a variable nested in the \sym{TLSSession} object that contains a pointer to a virtual method table (vtable) for dynamic dispatch.
By controlling the pointer to the object, the attacker controls the pointer to the virtual method table and gains full control over the target of an indirect call.

\begin{figure}[t]
\centering
\begin{lstlisting}[language=edl]
/* ECALL Definition in EDL */
public void* tls_client_new()
public int tls_client_write(
    [user_check]   void* session,
    [in, size=cnt] char* buf,
                   int cnt);
\end{lstlisting}
\begin{lstlisting}[language=rust]
// Rust Source Code 
pub extern "C" fn tls_client_write(
                session: *const c_void,
                bu: * const c_char,
                cnt: c_int)  -> c_int {
 @\circleone{}\,@ if session.is_null() {
        return -1;
    }

 @\circletwo{}\,@ if rsgx_raw_is_outside_enclave(
              session as * const u8,
              mem::size_of::<TlsClient>()) {
        return -1;
    }
    rsgx_lfence();

    let session = unsafe { &mut *(session as *mut TlsClient) };
\end{lstlisting}
\caption{Vulnerable Rust code: Check \circletwo can be bypassed.}
\label{fig:rust-code}
\end{figure}

The enclave code, as shown in Figure~\ref{fig:rust-code}, implements two pointer validation checks on the session pointer:
\begin{inparaenum}[(1)]
\item the pointer is checked to be not null \circleone and
\item not to be outside of the enclave \circletwo.
\end{inparaenum}
However, the check at \circletwo\ is not sufficient to protect the enclave since there are two possible bypasses.
First, the attacker can abuse a different \ecall to copy attacker-controlled data from the host application into the enclave memory (cf. \Cref{subsec:enclave:wolfssl}).
Second, the check at~\circletwo\ neglects that there are three memory states: outside, within the enclave, and partially inside the enclave.
Hence, \sym{outside\_enclave} and \sym{within\_enclave} are not strictly inverse, both return \sym{false} for any memory that is neither strictly outside nor strictly within the enclave.
The intention of the enclave developer for check~\circletwo was to assess whether the session pointer does indeed point to memory inside of the enclave, i.e.\ return an error if it is not strictly within (\mbox{\sym{if\,!\,rsgx\_raw\_is\_within\_enclave(\narrowellipsis)}} \mbox{\sym{return -1;}}).
This error belongs to the following pattern.

\begin{figure}[t]
  \begin{tabular}{lr}
    \begin{minipage}{0.6\linewidth}
      \begin{lstlisting}[language=C++]
sgx_is_outside_enclave(A, sz) == true
sgx_is_within_enclave(A, sz)  == false

sgx_is_outside_enclave(B, sz) == false
sgx_is_within_enclave(B, sz)  == true

sgx_is_outside_enclave(C, sz) == false
sgx_is_within_enclave(C, sz)  == false
\end{lstlisting}%
    \end{minipage} & \begin{minipage}{0.3\linewidth}
      \includegraphics[width=\linewidth]{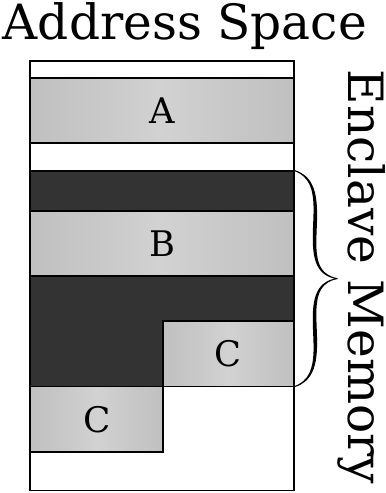}
    \end{minipage}
  \end{tabular}
  \caption{Possible buffer locations in SGX.}%
  \label{fig:overlapmemory}
\end{figure}

\pattern{\titlevulnOverlap}
For validating that an object is in secure memory, the \sgxsdk provides two functions: \sym{sgx\_is\_within\_enclave} and \sym{sgx\_is\_outside\_enclave}.
These functions check whether a memory area is \emph{strictly} outside or inside enclave memory.
However, they return unexpected results when handling edge-cases, where a memory buffer is overlapping both areas.
Figure~\ref{fig:overlapmemory} shows three different scenarios with buffers located either outside, inside, or outside as well as inside enclave memory. The validation functions from the \sgxsdk return \sym{false} for buffers that are overlapping both memory areas.

In the case of the \enclaveRustTLS, we can abuse the buggy check in \circletwo to bypass the pointer validation routine in our PoC exploit.
We allocate a page in the virtual address space right before the first page of enclave memory.
Thereafter, we place a fake \sym{TLSSession} object such that the last byte of the object is still part of enclave memory (i.e., the overlapping case).
This construction bypasses the validation at \circletwo since the memory is not strictly outside enclave memory.
However, the important part---the address of the vtable---is still stored in untrusted host memory.
Hence, we can fully control the target of an indirect jump and launch a code-reuse attack.
Our findings demonstrate that using a memory-safe language like Rust does not automatically ensure memory-safe enclaves. 
That is, the entire software stack must be guaranteed to be memory-safe.

The developers of the Rust SGX SDK acknowledged the problem and promptly updated their code.
Akin to our suggestions to the developers, the enclave code now utilizes session identifiers instead of pointers to identify TLS sessions; similar to using file descriptors on Unix-like systems.
Upon session creation in \sym{tls\_client\_new}, the pointer to the TLS session object is now inserted into a hashmap, which is then used to map the identifier in subsequent \ecalls.
Hence, no pointers are passed on the host-to-enclave boundary. This drastically reduces the attack surface of the enclave and eradicates both the vulnerability pattern \vulnResource and \vulnOverlap.

%% file: sections/enclaves/talos.tex
The open-source enclave \enclaveTalos supports terminating TLS inside of SGX enclaves within production webservers such as the Apache webserver~\cite{talos}.
To achieve this, \enclaveTalos introduces SGX specific patches to the \emph{libressl TLS implementation}.
The enclave exposes almost the entire TLS API of libressl over the \ecall interface, which utilizes many pointers that are marked as \verb+[user_check]+.
As such, this enclave contains the vulnerability patterns \vulnPtrInDS and \vulnResource.
However, the enclave does not simply return a raw pointer as it is the case for the enclaves \nameref{subsec:enclave:wolfssl} (\Cref{subsec:enclave:wolfssl}) and \nameref{subsec:enclave:rust} (\Cref{subsec:enclave:rust}).
Instead, it uses a shadowing mechanism that synchronizes selected fields (e.g.\ of the primary \sym{SSL} data structure) between the trusted and untrusted world.
This allows the host application to access some fields of the data structure, while keeping the actual copy in enclave memory~\cite{talos-report}.
This design choice was taken to allow unmodified web servers to interact with the SGX wrapped TLS API.
In principle, the shadowing mechanism is a legitimate pointer validation mechanism and allows the enclave to verify pointers passed by the untrusted host application.
However, the exposed API is quite comprehensive and \toolname discovered an \ecall that uses a function pointer in its data structure, where shadowing was missing.
This underlines the need for automated analysis tools, such as \toolname, to automatically identify missing pointer validation code.
Furthermore, we identified many potential sources for vulnerabilities in the code that handled the shadowing mechanism.
The shadowing mechanism failed to take into account that the \emph{NULL} pointer is a valid pointer in the SGX context.

\pattern{\titlevulnNull}
The special \emph{NULL} (or \emph{nullptr}) value is used in C/C++ code to signal that a pointer is not referencing any object.
However, it is represented by the numeric value~0, but on x86 systems (using virtual memory) the address~0 is a valid address.
Typically, there is no valid memory mapped to address~0. Hence, any accidental NULL pointer dereference results in a crash of the process (SEGFAULT).
However, a malicious host program or OS can map valid data at the page at address~0.
Thus, a NULL pointer dereference turns into a valid load and a bogus value from the page at address~0 is read instead of crashing the enclave.
This is similar to the kernel scenario, where the address~0 is typically a valid address in the user space.
As a mitigation, many OS kernels disallow mapping any memory at address~0.
However, for NULL pointer dereferences inside of SGX enclaves, there is currently no mitigation available, since the OS is considered untrusted in the SGX threat model.
As such, an enclave must assume that the page at address~0 is mapped into the address space.

Figure~\ref{fig:talos-code} shows the relevant code that contains a NULL-pointer dereference.
This snippet contains two mistakes: first, the pointer parameter \sym{out\_s} is supposed to point to the outside version of the TLS structure.
However, the enclave does not validate that the \sym{out\_s} actually points to outside enclave memory (\circleone).
As such, an attacker can simply pass some memory location inside of the enclave memory.
The function call at \circletwo retrieves the shadowed \sym{SSL} object that is within enclave memory.
However, when passing a bogus pointer this function will return a NULL-pointer to signal an error, which is not checked by the enclave.
The function call at \circlethree is the synchronization function that copies selected fields from the outside \sym{SSL} structure to the inside structure.
In case of an attack, the \sym{out\_s} pointer does point to an arbitrary location inside of the enclave, e.g., a secret key and \sym{in\_s} points to the NULL-page.
Thus, the enclave copies arbitrary data from enclave memory to the NULL-page resulting in an \emph{arbitrary read exploit}.

\begin{figure}[t]
\centering
\begin{lstlisting}[language=C,escapechar=@]
BIO* ecall_SSL_get_rbio(SSL *out_s) {
@\circleone{}\,@// out_s is not checked, can be in enclave memory
  /** Shadowing Mechanism **/
  hashmap* m = get_ssl_hardening();
  // returns NULL for invalid out_s
@\circletwo{}\,@SSL* in_s = hashmapGet(m, out_s);
  // copy arbitrary enclave memory to the NULL page
@\circlethree{}\,@SSL_copy_fields_to_in_struct(in_s, out_s);
@\circlefour{}\,@/* [...] libressl logic */
  // copy from the NULL page to arbitrary enclave memory
@\circlefive{}\,@SSL_copy_fields_to_out_struct(in_s, out_s); // [...]
\end{lstlisting}
\caption{Relevant parts of the EDL definition and C source code of the \enclaveTalos enclave.}
\label{fig:talos-code}
\end{figure}

Furthermore, the same bugs shown in Figure~\ref{fig:talos-code} can also be turned into an arbitrary write exploit primitive: for the function call marked with \circlefive, the enclave synchronizes back the fields of the inside structure to the outside copy.
In our NULL-pointer dereference attack, the variable \sym{in\_s} points to the NULL-page, while the variable \sym{out\_s} points to some arbitrary enclave memory location.
However, we have to overcome a race condition challenge to also control the value that is written.
Recall that the enclave first reads the value from enclave memory and thereafter writes the value to the NULL-page (\circlethree).
Hence, it would write back the same value to enclave memory that was copied to the NULL-page.
To tackle the race condition, we execute a different thread in the host application and change the contents of the NULL-page while the code between the two synchronization functions (\circlefour) is executed.
This effectively gives an attacker the arbitrary write capability.
Note that prior research has shown that it is trivial to win race conditions in the SGX threat model.
Since the attacker is in full control of the OS and the scheduling of the enclave's thread, the attacker can even single-step through the enclave code~\cite{VanBulck2017-sgxstep}.

\pattern{\titlevulnTOCTOU}
Enclaves run in an environment where it is easy to introduce Time-of-Check Time-of-Use (TOCTOU) bugs.
While the enclave developer can limit how many threads can concurrently enter an SGX enclave, the enclave developer has no control over the untrusted and possibly malicious OS.
When accessing host application memory, the enclave must assume that a separate host application thread can always change any content in the untrusted memory area.
As a consequence, an enclave cannot validate any data structures outside of the enclave memory.
In the \enclaveTalos example, we utilized a race condition similar to TOCTOU bugs to exploit the enclave.

%% file: sections/enclaves/synatee.tex
\Synaptics recently started to utilize SGX enclaves to securely process fingerprint data on Windows in \Lenovo and HP laptops.
The closed-source fingerprint driver contains a user space component with an SGX enclave.
\toolname discovered a control-flow hijacking primitive that can be exploited due to a NULL-pointer dereference (cf. \nameref{subsec:enclave:talos} in \Cref*{subsec:enclave:talos}).
The enclave utilizes a pointer in the global state, which is initialized as a NULL pointer.
Normally, this pointer would be initialized to point to a data structure inside enclave memory, but the attacker could potentially load the enclave and trigger the NULL pointer dereference without initializing this pointer.

We chose not to exploit the NULL-pointer dereference since the latest Windows versions strictly prohibit mapping a page at address~$0$.
That being said, the SGX threat model assumes that the attacker has full control over the OS, i.e., an attacker with OS privileges can disable this mitigation in the Windows kernel.
We demonstrated the feasibility of this in our PoC exploit for the \Goodix enclave (see \Cref{subsec:enclave:goodix}).
To avoid patching the Windows kernel and to make our PoC exploit more portable, we utilize a second finding of \toolname: a limited write exploit primitive due to an improperly sanitized pointer heavy data structure that is passed to the enclave.
This exploit primitive allows us to write a fixed byte-value to an arbitrary address.
We used this in our PoC exploit to first corrupt the pointer in the global state of the enclave to make it point to a fixed address in untrusted host application memory.
Next, we mapped our exploit payload to this fixed address thereby avoiding allocation of a page at address~$0$.

We chained two exploit primitives in our PoC Exploit, both discovered by \toolname.
The vulnerabilities we identified are due to the code patterns \vulnPtrInDS (cf. \nameref{subsec:enclave:intel} in \Cref*{subsec:enclave:intel}) and \vulnNull (cf. \nameref{subsec:enclave:talos} in \Cref*{subsec:enclave:talos}).

%% file: sections/enclaves/goodix.tex
\newcommand{\px}{$C_{16}$}
\newcommand{\py}{$F_{64}$}

The fingerprint reader driver is shipped on recent \Dell laptops and uses SGX enclaves to process biometric data.
The black-box analysis of \toolname discovered multiple limited controllable write primitives to arbitrary addresses.
For our exploit, we combined two of them to achieve a full control-flow hijack.

The first primitive, denoted as \px, discovered by \toolname copies a \SI{16}{\bit} value loaded from a NULL-pointer (see \Cref{subsec:enclave:talos}) to the address supplied in the \ecall argument by an attacker.
We patched the Windows kernel using a kernel debugger and disabled the check that prevents Windows user space applications to map the address~$0$, allowing us to exploit the NULL-pointer dereference in the enclave.
While the attacker controls the value and the address in the first primitive \px, due to the limited size of the controlled value, this primitive can only partially overwrite the instruction pointer.
Although this partial overwrite is often sufficient~\cite{phrack:partialeip}, we combine it with a second primitive also discovered by \toolname to achieve a full (64-bit) arbitrary write.
The second primitive, denoted as \py, is a limited write primitive that copies a \SI{64}{\bit} value loaded from a fixed address $A$ that is within secure memory to an attacker-controlled pointer in the \ecall argument.
We execute primitive \px\ four times to copy a full \SI{64}{\bit} value in \SI{16}{\bit} chunks to the address $A$, which is used in primitive \py.
This gives us control over the \SI{64}{\bit} value that is written by \py.
Subsequently, we can then use primitive \py\ to overwrite, e.g., a return address in secure memory.

The analysis of this enclave demonstrates that the vulnerability report produced by \toolname (cf. \Cref{sec:architecture}) provides sufficient information to easily create a PoC exploit for enclaves where source code is not available. 
We only needed to combine two primitives and for both \toolname reported the source and target addresses of the writes and the necessary \ecall arguments.

\let\px\undefined
\let\py\undefined

%% file: sections/disclosure.tex
We provided the developers of all the vulnerable enclaves a detailed report explaining the problematic code patterns, a working PoC exploit, and suggested fixes.
All of them confirmed our findings.
We supported the enclave developers by validating the patched versions with \toolname.
\Cref{tab:enclaveversions} shows the version number of the fixed enclave code, as far as they were available to us.
As a response to our report, \intel\ changed the code of the \enclaveIntelGMP\ enclave to use a serialization-based approach for parameters crossing the host-to-enclave boundary.
Since serialization avoids passing raw object pointers at the host-to-enclave boundary, the vulnerabilities were successfully fixed.
The developers of both, the \enclaveWolfSSL\ and the \enclaveRustTLS, followed our suggestions and stopped using pointers as resource references.
Both enclaves now utilize integer identifiers to look up the respective TLS session objects in a table inside of enclave memory.
The original developer of the \enclaveTalos\ acknowledged our findings, but notified us that he lacks the resources to develop fixes.
As such, this project must now be considered a deprecated and abandoned research project.
\Synaptics\ issued {CVE-2019-18619}~\cite{synaptics-cve} for the vulnerabilities we reported.
Given the high sensitivity of biometric data, they promptly developed a patch.
After coordinated disclosure with OEM vendors patches were published in July, 2020 (HP~\cite{hp-advisory}, Lenovo~\cite{lenovo-advisory}).
The security team of \Goodix developed a patch that we successfully verified with \toolname\ and \Goodix\ issued {CVE-2020-11667}.
As of July, 2020 patched drivers for Dell laptops are available~\cite{dell-advisory}.

%% file: sections/performance.tex
In this section, we analyze the efficiency and effectiveness of \toolname.
We focus our analysis on the three enclaves \enclaveIntelGMP, \enclaveRustTLS, and \enclaveWolfSSL since for these
\begin{inparaenum}[(1)]
  \item the source code is available, and
  \item a patched version already exists.
\end{inparaenum}
These insights allow us to compare \toolname' behavior on the vulnerable and fixed enclaves and reason about the occurrences of false alarms.

\subsection{Performance and Memory Usage}

\begin{figure}
\begin{center}
\hspace{.5em}%
\includegraphics[width=.8\linewidth]{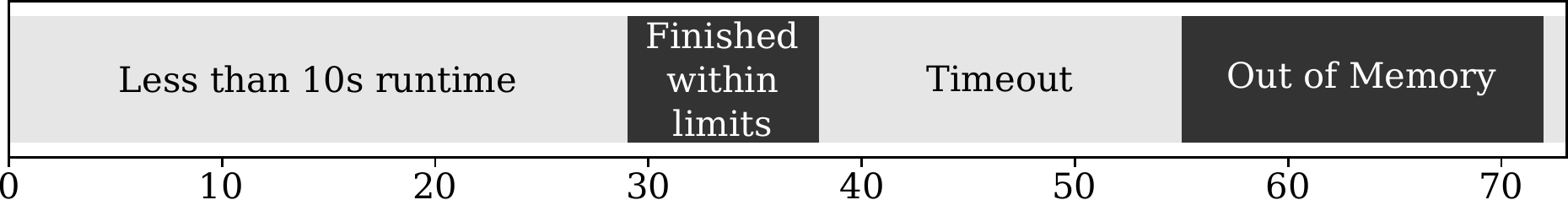}\\[2ex]
\includegraphics[width=\linewidth]{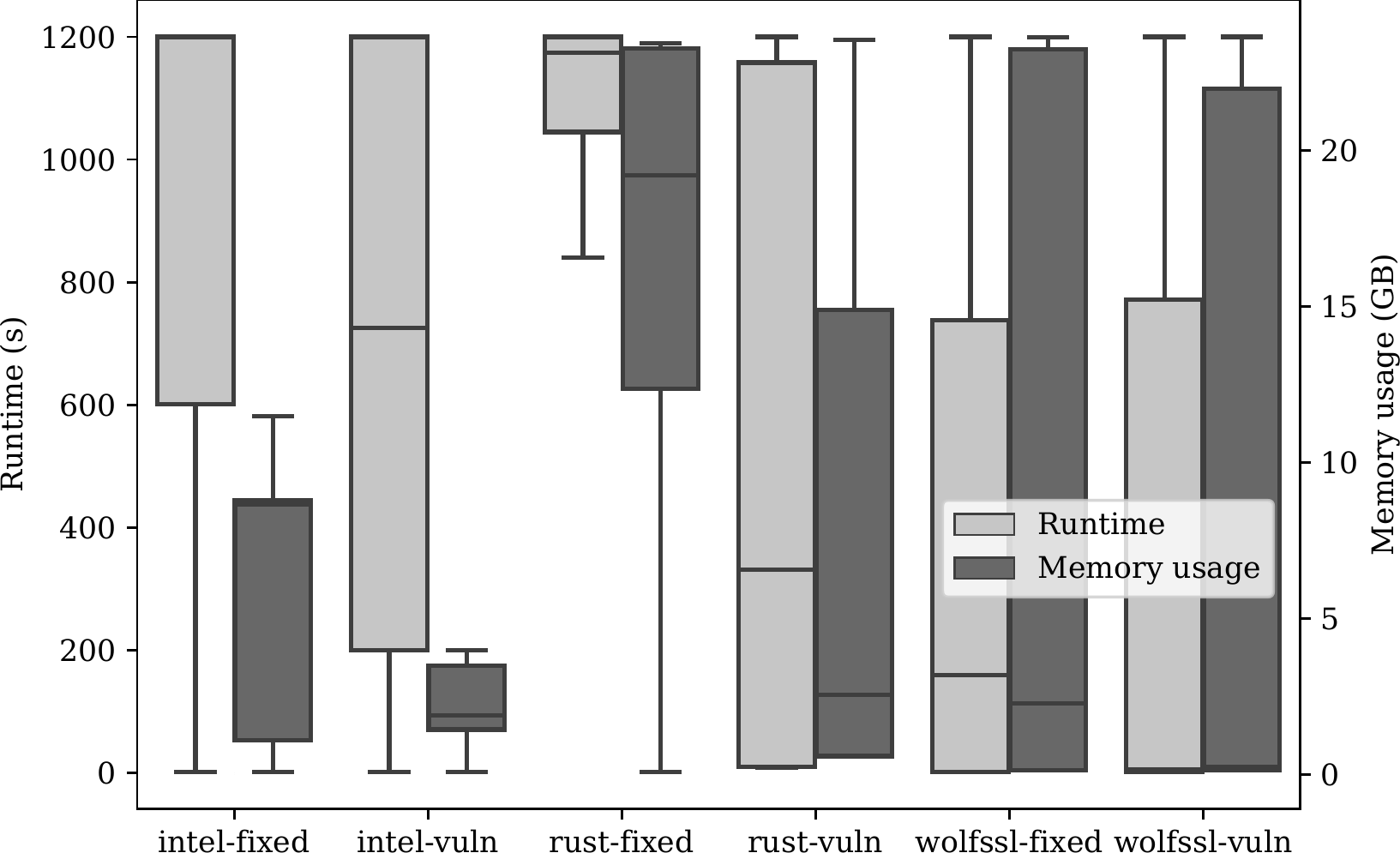}
\end{center}
\caption{Runtime and memory usage of the benchmarked enclaves.}%
\label{fig:rt-mem-boxplot}
\end{figure}

Our strategy is as follows: We analyze each \ecall using \toolname for a maximum of \SI{20}{\minute} using one CPU core up to a memory limit of \gb{24}.
The analysis was conducted on an AMD EPYC Processor with \SI{3.7}{\giga\hertz} and \gb{100} RAM allowing us to analyze up to 4~\ecalls in parallel.
\toolname utilizes angr version $8.20.1.7$ running on CPython~$3.6.9$ and Ubuntu~$18.04.4$.
All the exploitable primitives that we utilized in our PoC exploits are discovered within our time window of \SI{20}{\minute}.

For the three enclaves (\enclaveIntelGMP, \enclaveRustTLS, and \enclaveWolfSSL), we analyzed the 73\,\ecalls in detail.
The results are depicted in \Cref{fig:rt-mem-boxplot}:
the average memory usage over all \ecalls of those enclaves is \gb{8.8} ($\sigma = \gb{9.8}$).
The significant deviation for memory usage is mainly due to the highly variable size and complexity of the \ecalls.
Out of the analyzed \ecalls \p{40} finished within \s{10},
\p{52} finished within the given limits, and \p{48} exceeded the limits (\p{23} by time, also \p{23} by memory, and \p{1} by time and memory).

Our analysis in Section~\ref{sec:evaluation} demonstrates that using this analysis strategy is sufficient to successfully uncover problematic code patterns.
While symbolic execution is a powerful analysis technique, it requires high computing resources (CPU time and memory) to explore the state space of a program.
Hence, it is only natural that the analysis of some of the \ecalls hits the resource limits we defined for the benchmarking experiments. However, during a security analysis of an enclave, the analyst can schedule more time and memory as needed for specific \ecalls.
Furthermore, we did not yet implement all of the advanced techniques to improve the efficiency of symbolic execution that was proposed in prior work~\cite{Baldoni2018symexsurvey,Avgerinos2016veritesting,Cha2012mayhem}. This was simply not necessary to discover vulnerabilities in our set of analyzed enclaves.

\subsection{Accuracy and False Alarms}

Since the analysis of \toolname does focus on soundness rather than on completeness, the number of false alarms is rather small.
Note that a complete false positive analysis is impossible as we lack any ground-truth, i.e., all of our findings are zero-day vulnerabilities and we are unable to provide any comparison of \toolname to related approaches since there does not yet exist any other automated vulnerability discovery approach for SGX enclaves.
Hence, we opted for the following strategy: we confirmed \toolname' alarms by constructing PoC exploits and disclosing our findings to the affected vendors.
After the vendors fixed the vulnerabilities, we manually verified that the enclaves' source code (for \enclaveIntelGMP, \enclaveRustTLS, and \enclaveWolfSSL) does not contain further vulnerabilities.
These patched enclaves give us a limited form of ground-truth as any finding in the updated enclaves is a false alarm.

In our analysis of the three vulnerable enclaves \toolname produced 149 findings.
By constructing a proof-of-concept exploit based on the findings of \toolname, we confirm that those findings were indeed true alarms.
We selected gadgets in shallow program paths containing the least conditions on the initial state (i.e., constraints on the enclave's pre-\ecall state) and then constructed a PoC exploit based on the selected gadgets.

Thereafter, we analyzed the patched versions of the enclaves.
\toolname confirmed that our original and exploited findings are not longer present in the patched enclaves. However, the analysis of \toolname still produced 56 findings.
Our root-cause analysis of those findings reveals a possible indicator of a false alarm in \toolname' reports: global memory is treated as unconstrained symbolic value by \toolname (see challenge C4 in \Cref{sec:impl-challenges}).
For example, the patched \enclaveIntelGMP utilizes an initializing \ecall which sets up a function pointer in global memory.
\toolname discovered that other \ecalls do not check that function pointer before use.
Due to the \ecall-centric analysis of \toolname, the function pointer is considered unconstrained and a controlled jump is reported.
However, in reality, the function pointer can only take fixed values.
Thus, this finding on its own is not exploitable.
On the other hand, in case \toolname\ would have also discovered a controlled write primitive, we still would have been able to construct a proof-of-concept exploit.
The false positives that we encountered in the other patched enclaves (\enclaveRustTLS and \enclaveWolfSSL) are caused by the same issue.
In our future work, we plan to annotate and filter such false alarms as \emph{low severity} based on \toolname' pointer-tracking component.

%% file: sections/discussion.tex
\paragraph{Analyzing \ocalls.}
\toolname puts its focus on \ecalls as those are the prevalent way to pass data to enclaves. Further, since \ocalls are only reachable through \ecalls, their support is a precondition for \ocalls.
Nevertheless, we plan to implement \ocall-support in our future work.

Handling the \ocall interface is particularly challenging due to the lack of semantic information.
From a binary analysis point-of-view, an \ocall is not easily distinguished from a regular return from an \ecall, i.e., both utilize the \emph{EEXIT} instruction to exit the enclave.
As such, \toolname will stop executing a program path in the enclave once an \ocall (or \emph{EEXIT}) is reached and thus will not analyze any \ecall code beyond the first \ocall.
To overcome this limitation, we utilize symbol information to detect \ocall invocations in \toolname.
If \toolname discovers that an \ocall is executed, e.g., due to symbols and functions of the \sgxsdk, then \toolname will skip the execution of the \ocall and set the return value of the \ocall to an unconstrained symbolic value.
This allows \toolname to continue the analysis after the \ocall with a rough over-approximation of the \ocall's effects since the actual semantics of the \ocall are not emulated.
We leave the development of a heuristic to detect \ocalls on a binary-level without symbols as future work.

\paragraph{Manual Effort with \toolname.}
\toolname automatically detects vulnerabilities in enclaves.
More specifically, \toolname reports the exploit primitives resulting from the vulnerabilities.
For instance, \toolname will show the location of a controlled write combined with the constraints (i.e., possible values) on the address, value, and path that leads to the write instruction.
An analyst must then inspect the report and decide whether the findings or any combination of findings is exploitable, or if the alarm is a false positive.
While the information reported by \toolname is sufficient to construct PoC exploits, we plan to incorporate exploit generation schemes as proposed in prior work~\cite{Avgerinos2011aeg,Cha2012mayhem,Hu2015flowstitch,heelan2019gollum} into \toolname to automatically synthesize a malicious host application that reproduces the crash.

\paragraph{Fuzzing Enclaves.}
Coverage-guided fuzzing is another prominent technique to identify vulnerabilities in binary code~\cite{afl}.
In contrast to symbolic execution, fuzzing scales well to large software projects. 
As such, fuzzing would potentially allow analysis of large and complex enclave binaries to tackle the general problem of path explosion. On the other hand, applying fuzzing to SGX enclaves is not straightforward:
\begin{inparaenum}[(1)]
  \item To ensure efficiency, fuzzing requires a sophisticated mutation strategy. However, mutation for the complex \ecall interface requires significant engineering effort.
  \item Fuzzing relies on dynamic analysis tools to instrument binaries~\cite{pin,dyninst}, which are currently not available for SGX enclaves.
\end{inparaenum}
In particular, integrating dynamic analysis tools is highly challenging when analyzing proprietary enclave binaries. 
Note that the protection mechanisms provided by SGX impede dynamic binary instrumentation.
Further, static binary instrumentation often fails to accurately rewrite binaries.
Consequently, we decided to rely on symbolic execution as it allows us to fully control the simulated environment.
Further, it comes with additional flexibility significantly simplifying implementation and integration of symbolic vulnerability detectors.
However, enabling hybrid fuzzing/concolic execution in \toolname\ is worthwhile investigating for future work.

%% file: sections/relatedwork.tex
The security research on privilege separation lead to system architectures that separate user from kernel space.
However, several kernel vulnerabilities bypassed this separation simply because the kernel is not strictly separated from user space~\cite{Chen2011-fb,Kemerlis2012kguard,Gens2018-jn,freebsdnullptr,linuxnullptr}.
As a response, CPU vendors introduced hardware-based mitigation mechanisms, such as SMAP or SMEP~\cite{IntelArchManualVol3}, to enforce stricter separation.
In fact, there are many parallels between the user/kernel space interface and the SGX host-to-enclave interface.
That is, a higher privileged partition (the enclave) must carefully parse and validate any data that is written by the untrusted partition (the host application).

Prior work in this area introduced mechanism allowing a user space program to reliably execute in the presence of a compromised operating system~\cite{Lie2003untrustedos,Ports2008-rb,Chen2008overshadow,mccune2008flicker}.
However, \citewauthor{checkoway2013iago} have shown that existing legacy software cannot be simply retrofitted to such environments mainly because many kernel and operating system APIs implicitly assume that the kernel is the most trusted part of the system, e.g., in the threat model of a traditional Unix-like system the kernel is assumed to have full control over the code and data areas of any user space process.
As such, existing software, such as most implementations of the C standard library, lack any validation of data passed from the kernel.
So-called \emph{Iago} attacks exploit this fact and show that a malicious kernel can easily corrupt memory of a user space process by returning bogus arguments from system calls.
As we show in this paper, very similar issues apply to SGX enclaves; especially when legacy code is retrofitted to run inside SGX enclaves.

\citewauthor{Hu2015-privsepvulns} showed that any software that is separated into equally-privileged but mutually untrusted partitions can be vulnerable to similar attacks.
They presented an approach based on taint tracking and constraint solving to detect arbitrary write and possible TOCTOU vulnerabilities for a limited number of execution paths.
In contrast, \toolname utilizes full symbolic execution to identify arbitrary write primitives.
Furthermore, \toolname also discovers control-flow hijacking and NULL-pointer dereferences.
\toolname' analysis also includes scenarios, where the exploit depends on the global state of the target.

Recently, \citewauthor{vanbulck2019twoworlds} presented a security analysis of several TEE SDKs, whereas we focus on analyzing enclaves.
They identified vulnerabilities in TEE SDKs using only manual code review.
In contrast, we introduce an \emph{automated} vulnerability detection framework for SGX enclave binaries, which additionally assists an analyst in assessing the vulnerability and constructing an exploit.

Many Android phones using ARM processors utilize the TrustZone trusted execution environment (TEE) to protect critical software.
In contrast to SGX, TrustZone splits all privilege levels into a trusted and untrusted world, where the trusted OS has the highest privilege on the system.
\citewauthor{machiry2012boomerang} analyzed the attack surface of the privilege boundary between normal world and TEE.
They identified a class of vulnerabilities caused by to the semantic gap between normal world and TEE.
They allow unprivileged, untrusted user space applications (e.g., a sandboxed Android app) to abuse the TEE to compromise the normal OS (the Linux kernel).
This type of vulnerability does not apply to SGX as enclaves have little privileges and are prohibited to interact with the OS.
\citewauthor{harrison2019partemu} implemented a fuzzer based on full-system emulation of the TrustZone TEE including the trusted OS and trusted applications.
The main challenge for analyzing ARM-based TEEs is the fact that a custom trusted OS, including required hardware, must be emulated.
In contrast, SGX enclaves generally lack direct hardware access. 
Further, as discussed in \Cref{sec:discussion}, symbolic execution offers several advantages over fuzzing when analyzing SGX enclaves.

%% file: sections/conclusion.tex
Intel SGX is a promising security technology to strongly isolate sensitive code and data into enclaves. 
However, implementing the host-to-enclave boundary securely is highly critical as the enclave processes and operates on input originating from untrusted memory space. 
To allow thorough security testing of this interface, we perform a systematic investigation on publicly available SGX enclaves. 
A major contribution of this paper is to introduce an automated analysis approach to determine vulnerabilities in enclaves. 
To do so, our approach develops a sophisticated symbolic execution framework that is able to analyze enclave binaries and produce detailed vulnerability reports to significantly simplify the construction of proof-of-concept (PoC) exploits. 
Our findings on public enclaves reveal vulnerabilities in two fingerprint drivers (by \Synaptics and by \Goodix), three TLS libraries, and a project published by \intel. 
For each, we constructed PoC exploits to confirm the severity of the vulnerability and perform control-flow hijacking allowing an attacker to subvert any confidentiality or integrity guarantees offered by the SGX enclaves.
We analyzed the root causes of the vulnerabilities and identified vulnerability patterns that likely also affect privately deployed enclaves.
Addressing our findings is crucial to allow secure deployment of SGX enclaves.